\documentclass[conference]{IEEEtran}
\usepackage{graphicx}
\usepackage{verbatim}
\usepackage{float}

\makeatletter
\def\footnoterule{\relax%
  \kern-5pt
  \hbox to \columnwidth{\hfill\vrule width 0.8\columnwidth height 0.4pt\hfill}
  \kern4.6pt}
\makeatother

\usepackage{multirow}
\usepackage{dblfloatfix}

\IEEEoverridecommandlockouts

\IEEEaftertitletext{\vspace{-2\baselineskip}}
\begin{document}

\title{Near-Landauer Reversible Skyrmion Logic with Voltage-Based Propagation\vspace{-10 px}}
\author{
\IEEEauthorblockN{
    Benjamin W. Walker\IEEEauthorrefmark{1},
    Alexander J. Edwards,\IEEEauthorrefmark{1}\IEEEauthorrefmark{2}
    Xuan Hu\IEEEauthorrefmark{1},\\
    Michael P. Frank\IEEEauthorrefmark{2},
    Felipe Garcia-Sanchez\IEEEauthorrefmark{3},    
    Joseph S. Friedman\IEEEauthorrefmark{1}\vspace{0 px}
}

\IEEEauthorblockA{
    \IEEEauthorrefmark{1}University of Texas at Dallas, Richardson, TX;
    \IEEEauthorrefmark{2}Sandia National Laboratories, Albuquerque, NM;\\
    \IEEEauthorrefmark{3}Universidad de Salamanca, Salamanca, Spain\vspace{10 px}
}
}

\maketitle
\begin{abstract}
Magnetic skyrmions are topological quasiparticles whose non-volatility, detectability, and mobility make them exciting candidates for low-energy computing. Previous works have demonstrated the feasibility and efficiency of current-driven skyrmions in cascaded logic structures inspired by reversible computing. As skyrmions can be propelled through the voltage-controlled magnetic anisotropy (VCMA) effect with much greater efficiency, this work proposes a VCMA-based skyrmion propagation mechanism that drastically reduces energy dissipation. Additionally, we demonstrate the functionality of skyrmion logic gates enabled by our novel voltage-based propagation and estimate its energy efficiency relative to other logic schemes. The minimum dissipation of this VCMA-driven magnetic skyrmion logic at 0~K is found to be $\sim$6$\times$ the room-temperature Landauer limit, indicating the potential for sub-Landauer dissipation through further engineering.
\end{abstract}
\renewcommand\IEEEkeywordsname{Keywords}
\begin{IEEEkeywords}
reversible computing, conservative logic, spintronics, magnetic skyrmion, VCMA, energy-efficient computation
\end{IEEEkeywords}
\section{Introduction}
\bstctlcite{IEEEexample:BSTcontrol}

The non-volatility and energy-efficient mobility of magnetic skyrmions have made them promising candidates for computation. Previous work has utilized skyrmions in a variety of logical devices \cite{HeydermanAPL2021,Zhang2019,SZhang2015,XZhang2015,XZhang2015(2),xing2016,Luo2018,XZhang2015(3)}, and recently skyrmions have emerged as exciting candidates for reversible computing \cite{ChauwinPRA2019, WalkerAPL2021, HuIEEE2022}. By conserving information and maintaining logical reversibility in an adiabatic manner, these systems avoid the limits on thermodynamic efficiency intrinsic to traditional computing schemes \cite{FredkinIJTP1982}. Therefore, such reversible computing systems have the potential to outperform the $kT \: \ln(2)$ limit determined by Landauer \cite{LandauerIBM1961}. 

The reversible skyrmion logic system of \cite{ChauwinPRA2019, WalkerAPL2021, HuIEEE2022} uses a heavy metal/ferromagnet heterostructure to allow for skyrmion stability within the ferromagnet via the Dzyaloshinsky–Moriya interaction. The spin-Hall effect induces skyrmion propagation with applied electronic current, while the skyrmion-Hall effect and skyrmion-skyrmion repulsion produce billiard-ball-like interactions within the logic gates. Skyrmions can be synchronized through voltage-controlled magnetic anisotropy (VCMA) by modulating the perpendicular magnetic anisotropy (PMA) with the application of a voltage on an electrode \cite{WalkerAPL2021}. 

While adiabatic reversible CMOS computing has been well-studied \cite{FrankIEEE2020}, reversible skyrmion logic gets closer to the Landauer limit by directly implementing the elastic billiard ball model proposed by Fredkin and Toffoli \cite{FredkinIJTP1982}. However, the electrical current required by \cite{ChauwinPRA2019, WalkerAPL2021, HuIEEE2022} dissipates significant energy. This work therefore proposes a new method of skyrmion propagation that uses voltage-controlled magnetic anisotropy (VCMA) to eliminate the need for current-driven propagation, enabling reversible skyrmion computing to achieve near-Landauer energy dissipation.

\section{VCMA-Driven Skyrmion Propagation}
\begin{figure}[tb]
    \centering
    \includegraphics[width=1\columnwidth]{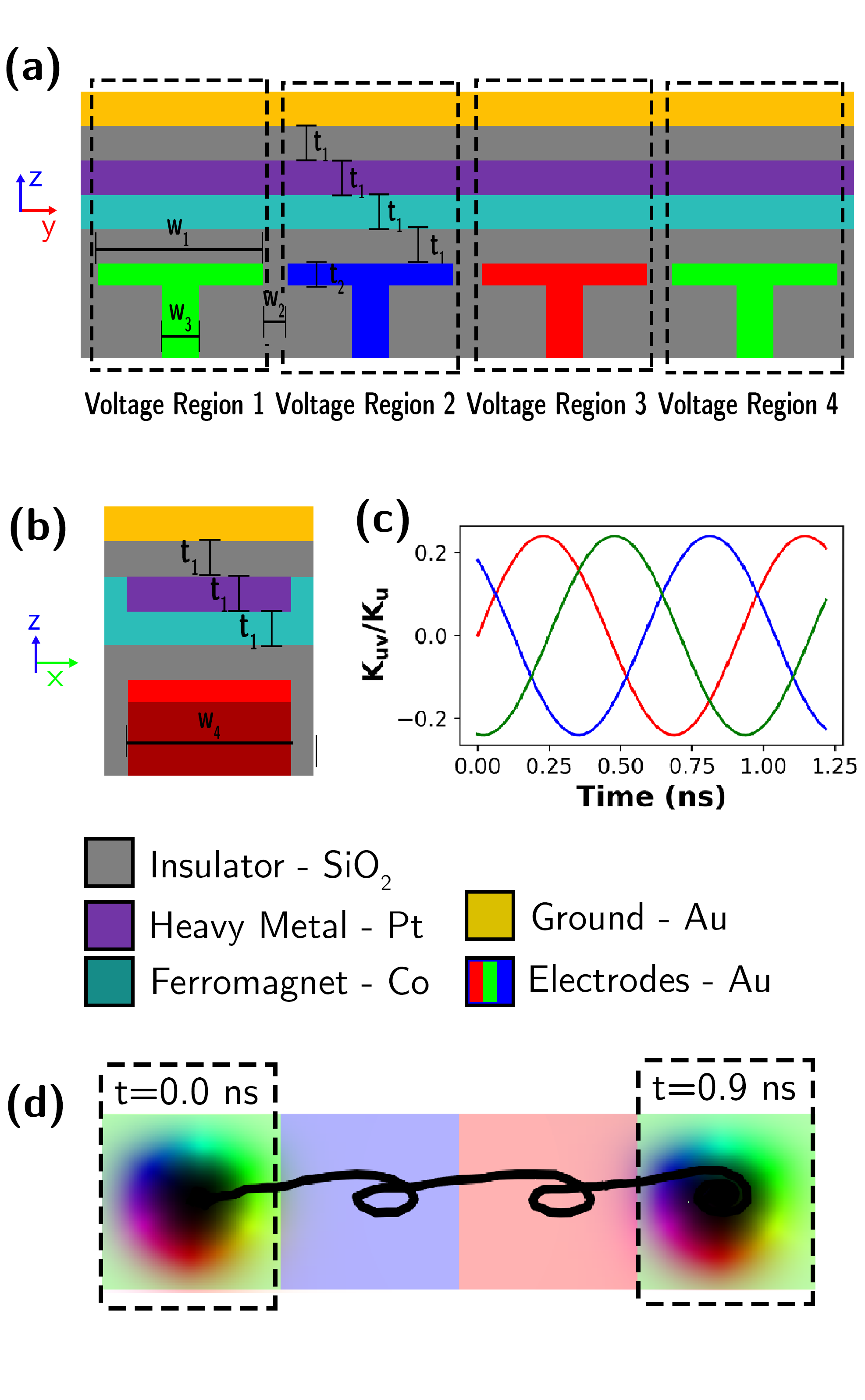}
    \caption{\vspace{0 px}Skyrmion propagation methodology. An electric field is applied across a heavy metal/ferromagnet interface (teal/purple) via electrodes (red/green/blue) separated by an insulating dielectric (grey). The skyrmion exists in the ferromagnet layer. (a) YZ Device Cross Section (b) XZ Device Cross Section ($w_1 =  19$nm, $w_2 =  1$nm, $w_3 =  11$nm, $w_4 =  20$nm, $t_1 =  0.8$nm, $t_2 =  0.5$nm) (c) VCMA anisotropy waveform applied to electrodes of corresponding color. (d) Micromagnetic simulation of skyrmion propagation; black path indicates skyrmion trajectory.}
    \label{fig:structure}
\end{figure}

We propose extremely efficient skyrmion propagation with a three-phase sinusoidal voltage that modulates magnetic anisotropy through VCMA, driving the skyrmions towards regions of lower anisotropy without requiring electrical current \cite{LiuPRA2019}. Though previous works modulate PMA periodically in discrete steps \cite{LiuPRA2019}, the rapid charging of electrode capacitance results in significant energy dissipation. By using three voltage sinusoids each shifted by $\frac{2\pi}{3}$ radians applied to neighboring electrodes (Fig. \ref{fig:structure}), skyrmions can be controllably propagated.

\begin{figure}[tb]
    \centering
    \includegraphics[width=1\columnwidth]{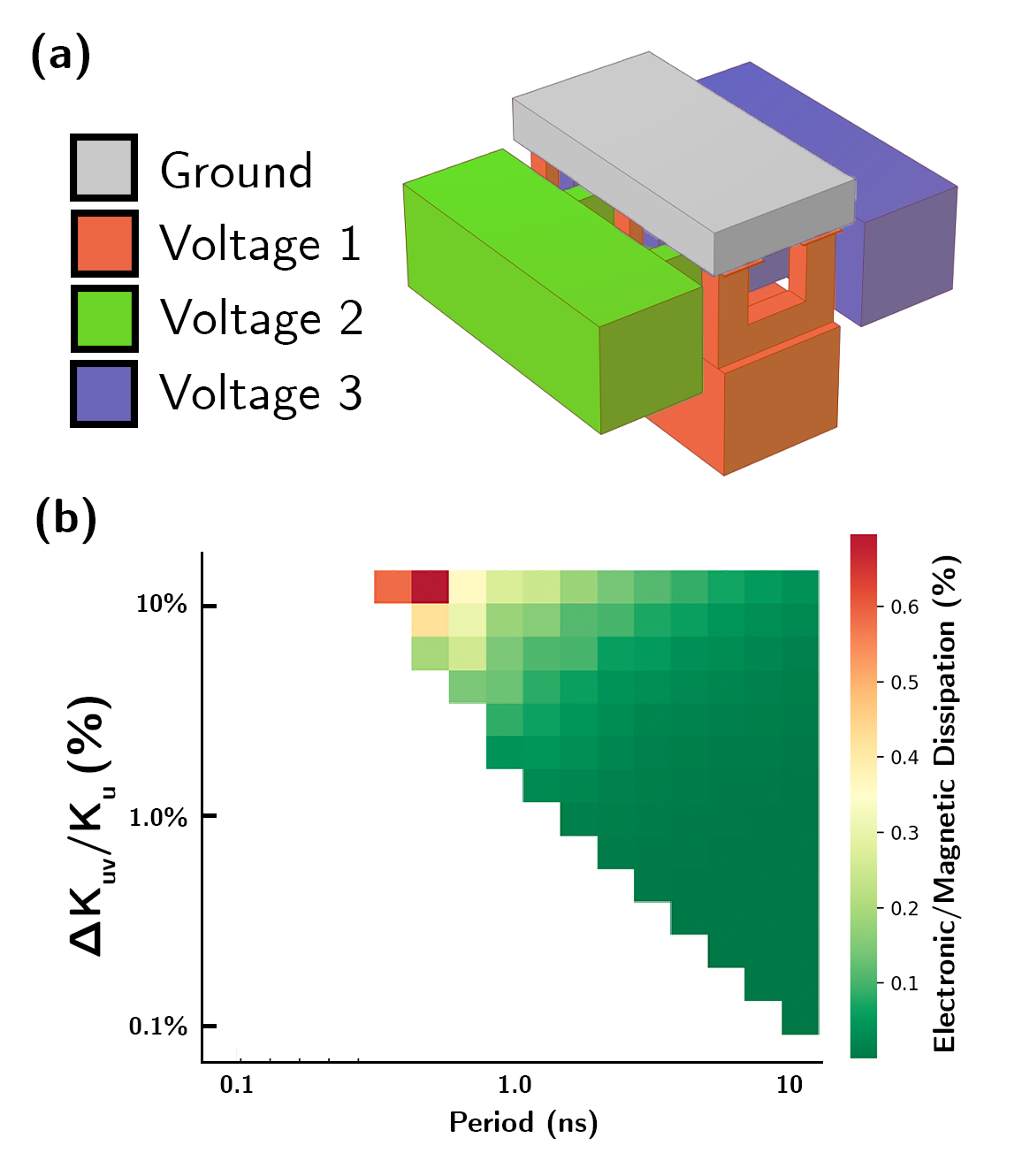}
    \caption{(a) Three-dimensional electrical model. Sinusoidal voltage waveforms applied to the external face of each volume. The resistivity and self-capacitance of the device determine the electronic dissipation. (b) Ratio of magnetic and electronic dissipation for a skyrmion wire.\vspace{-15 px}}
    \label{fig:3d_device}
\end{figure}

The magnetic dissipation of this VCMA-driven propagation scheme can be calculated \cite{VotoThesis2017} as $$\frac{dE}{dt} = -\frac{\alpha \mu_0}{\gamma_0 M_s}\int_V \left(\frac{d\bf{M}}{dt}\right) ^2 dV.$$ By implementing this equation in mumax3 \cite{Mumax}, the magnetic dissipation was calculated for all simulations. A 3D model for the AND/OR device was modeled using COMSOL (Fig. \ref{fig:3d_device}); for a range of temperatures, frequencies, and voltages, the Ohmic losses of the device were calculated to estimate electronic dissipation. As shown in Fig. \ref{fig:3d_device}, the magnetic dissipation dominates the electronic dissipation. Therefore, optimizing magnetic operation is more important than optimizing electronic parameters.

\begin{figure}[tb]
    \centering
    \includegraphics[width=1\columnwidth]{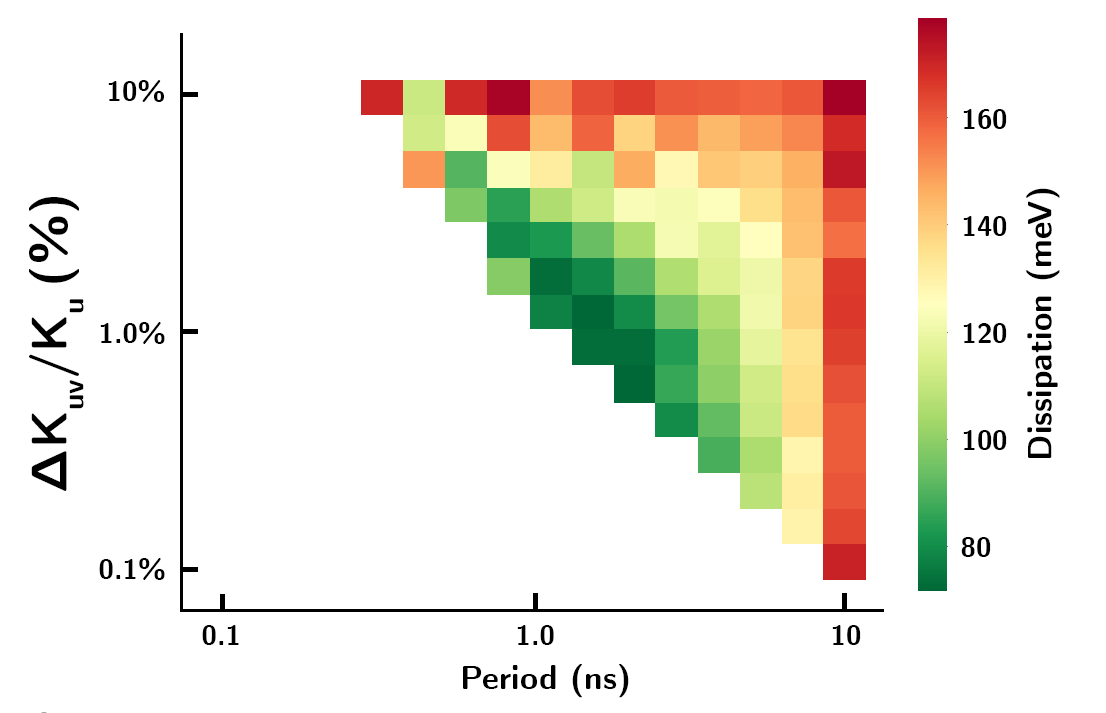}
    \caption{Dissipation heatmap for a skyrmion in a 60 nm long wire. Each non-white square represents simulation parameters where the skyrmion propagated correctly and in-sync. $\Delta K_{uv}/K_v$ is the maximum relative change in PMA between when voltage is applied and the baseline anisotropy. The color for each square represents the average dissipation (micromagnetic and electric) for the parameter combination.\vspace{-15 px}}
    \label{fig:wire_map}
\end{figure}

\begin{figure*}[!htb]
    \centering
    \includegraphics[width=\textwidth]{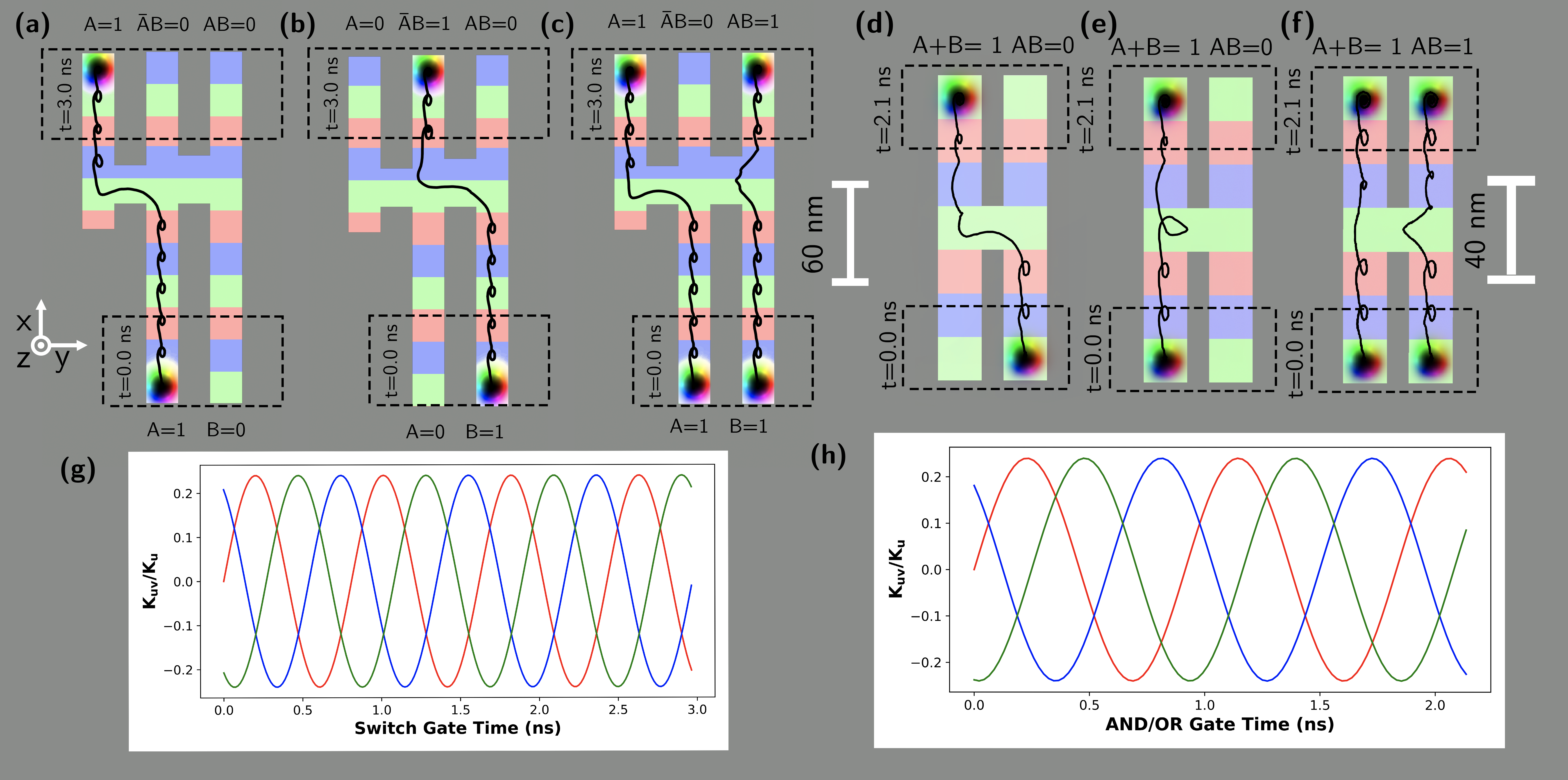}
    \caption{Micromagnetic simulation results for Ressler-Feynman Switch and AND/OR gate using voltage-based propagation. Skyrmions (colored circles) move in +y as indicated by their trajectory in black. Each colored region represents a VCMA voltage zone, each of which are $\frac{2\pi}{3}$ radians out of phase. Simulations shown for switch gate with input combinations (a) A=0, B=1; (b) A=1, B=0, (c) A=B=1. Simulations shown for AND/OR gate with input combinations (d) A=0, B=1; (e) A=1, B=0, (f) A=B=1. Anisotropy waveform applied to each VCMA region for (g) switch gate and (h) AND/OR gate, where $\Delta K_{uv}/K_v$ is the relative change in PMA between when voltage is applied and the baseline anisotropy.}
    \label{fig:AndOr}
\end{figure*}
\begin{figure}[!htb]
    \centering
    \includegraphics[width=1\columnwidth]{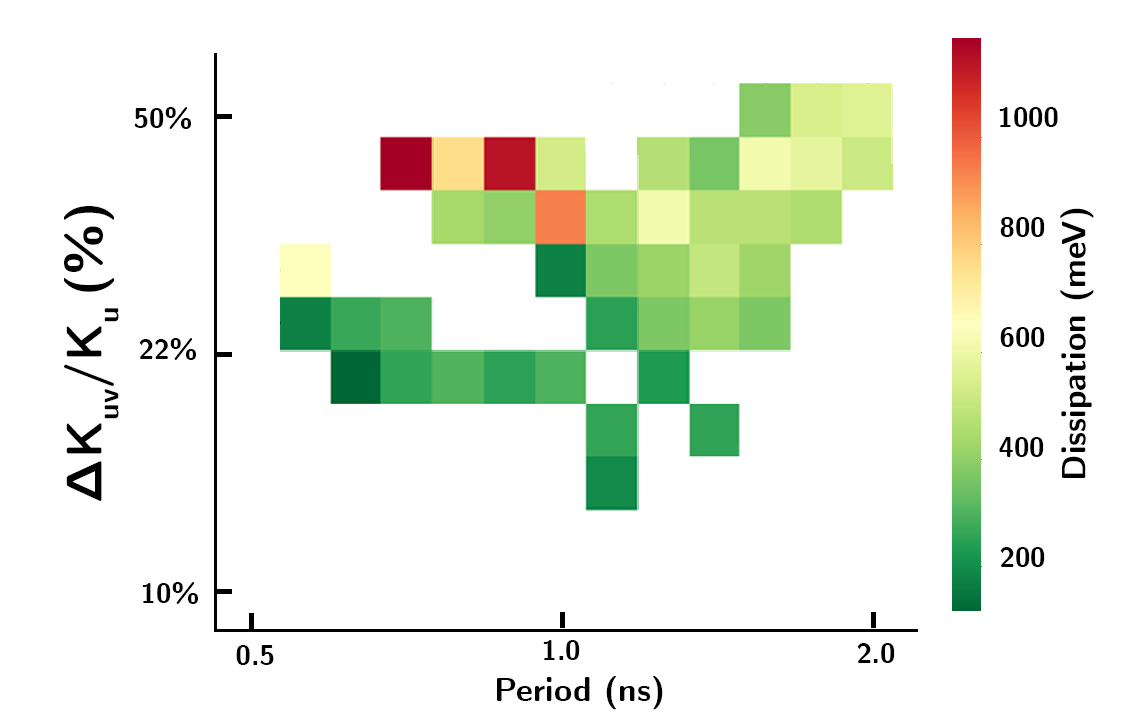}
    \caption{Dissipation heatmap for a skyrmion AND/OR gate. Each non-white square represents simulation parameters where the logic performed correctly for all input combinations. $\Delta K_{uv}/K_v$ is the maximum relative change in PMA between when voltage is applied and the baseline anisotropy. The color for each square represents the average dissipation (micromagnetic and electric) across all input combinations.\vspace{25 px}}
    \label{fig:logic_map}
\end{figure}
\begin{figure}[!htb]
    \centering
    \includegraphics[width=\columnwidth]{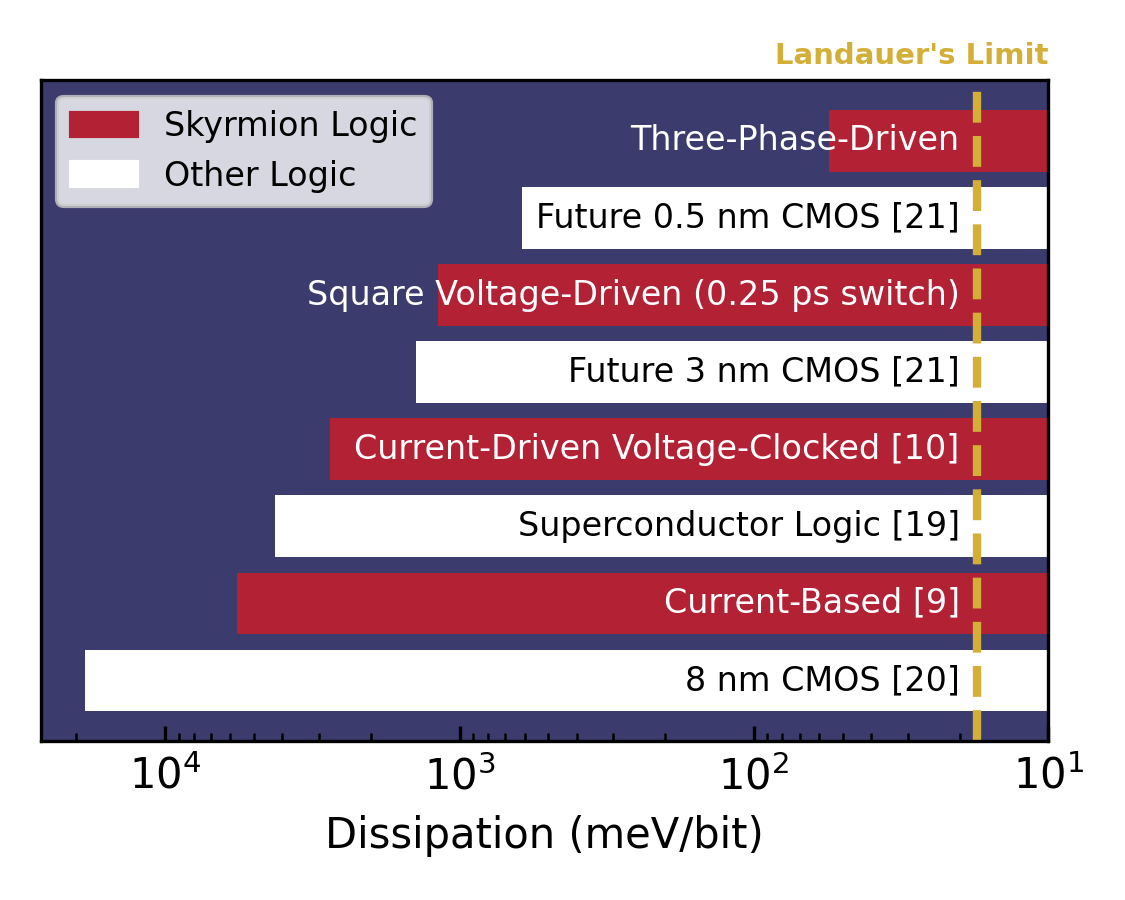}
    \caption{A comparison of total dissipation per bit processed for skyrmion logic technologies (red) and other logic technologies (white).  While temperature is not considered, considerable progress has been made in the efficiency of skyrmion-based logic as it approaches the room-temperature Landauer limit (gold line).}
    \label{fig:Comparison}
\end{figure}

Fig. \ref{fig:wire_map} shows the relationship between circuit parameters and the functionality and efficiency of the propagation scheme. Because skyrmion acceleration increases with PMA gradient, large anisotropy increases result in shorter acceleration times. Therefore, as the effective anisotropy is increased, the period can be decreased while allowing for proper propagation. While electronic dissipation is inversely proportional to the square of period, the micromagnetic dissipation dominates the dissipation, being proportional to the period. Further, both electronic and magnetic dissipation increase with increased PMA strength. Therefore, there should be an optimal value where both PMA and period are minimized such that the circuit operates with minimum total dissipation. Fig. \ref{fig:wire_map} shows this section in dark green.

At this optimal value, the skyrmion wire dissipates 80 meV per 60 nm, for a unit dissipation of 0.214 pJ/m. This represents a decrease in electrical energy consumption of $\sim$8,000$\times$ relative to the square wave clocking of \cite{LiuPRA2019} and $\sim$50$\times$ compared to current-driven propagation \cite{WalkerAPL2021}. Further, this scheme has a throughput of 1 Gbps at peak efficiency, for a 1,800$\times$ increase in data transfer efficiency (Mbps/W) compared to \cite{Chen2022}.

\section{Voltage-Based Reversible Logic}
By integrating this highly-efficient VCMA-driven propagation into the reversible logic scheme of \cite{ChauwinPRA2019}, the energy dissipation is far smaller than that required with current-driven propagation. Fig. \ref{fig:AndOr} shows the operation of a Ressler-Feynman switch gate and an AND/OR gate, where three clock cycles bring the skyrmions to their respective outputs for the switch gate and two are required for the AND/OR switch gate. This VCMA-driven system circumvents the need for synchronizers required in \cite{ChauwinPRA2019, WalkerAPL2021}, as skyrmion position is directly controlled by the propagation clock.

The AND/OR gate is logically reversible only for the subset of input combinations with two or zero skyrmions; with one skyrmion, it is impossible to recover the inputs. However, the Ressler-Feynman gate is fully logically reversible for all input combinations. As shown in \cite{HuIEEE2022}, the reverse Ressler-Feynman gate is also logically reversible for the subset of outputs produced by the forwards Ressler-Feynman gate. However, the Ressler-Feynman gate remains physically irreversible, as the reverse trajectories differ from the forward ones. This hysteresis creates entropy and increases dissipation, preventing sub-Landauer computation.

As shown in Fig. \ref{fig:logic_map}, the heatmap for the AND/OR gate is more complicated than that of the skyrmion wire. Rather than a linear relationship between period and PMA, the AND/OR gate is far more complex. Firstly, the period cannot be larger than 2 ns, as the skyrmion will not have enough velocity to exhibit the skyrmion-Hall force to interact in the logic gate. The dependence on the skyrmion-Hall force prevents arbitrarily slow computation. Next, there are a few ideal bands where the period and PMA strength correlate for proper logical operation. Similarly to the skyrmion wire, magnetic dissipation dominates, where increasing period and PMA strength results in increased dissipation.

The system can run from $\sim$100 MHz to $\sim$1GHz, with minimum average dissipation of 102 meV/operation. While the micromagnetic simulations were performed at 0~K, the minimum dissipation is $\sim$6$\times$ the Landauer limit at room temperature. As the skyrmion dynamics are not physically reversible, this dissipation remains above the minimum predicted by Landauer. As illustrated in Fig. \ref{fig:Comparison}, this is $\sim$9,000$\times$ lower than 8 nm CMOS and $\sim$80$\times$ lower than superconductive logic \cite{Herr2011}, demonstrating the viability of voltage-based skyrmion logic as a low-energy computation scheme.

\section{Conclusions}

The proposed three-phase VCMA-driven skyrmion propagation is far more efficient than previous alternatives. Implementing this propagation mechanism into the reversible skyrmion computing system results in a dissipation only $\approx 6\times$ that of the Landauer limit. This represents a massive step forward in skyrmion logic efficiency.

The major obstacles preventing the scheme from computing below the Landauer limit are the lack of physical reversibility and insufficient adiabaticity. The logically reversible Ressler-Feynman Switch gate is not physically reversible, as the return paths of the outputs are not identical to that of the inputs. This results in hysteresis and the necessary generation of entropy, wholly preventing sub-Landauer dissipation. Further, there are inefficiencies such as the loops in the skyrmion trajectories which worsen adiabaticity. Once true physical reversibility is achieved, improvements to efficiency could result in sub-Landauer computation with skyrmion logic.

\section*{Acknowledgements}

This work is supported in part by the Advanced Simulation and Computing (ASC) program at the U.S. Department of Energy’s National Nuclear Security Administration (NNSA). Sandia National Laboratories is a multi-mission laboratory managed and operated by National Technology and Engineering Solutions of Sandia, LLC, a wholly owned subsidiary of Honeywell International, Inc., for NNSA under contract DE-NA0003525. This document describes objective technical results and analysis. Any subjective views or opinions that might be expressed in this document do not necessarily represent the views of the U.S. Department of Energy or the United States Government.  Approved for public release, SAND2023-11355O.

This research is supported in part by the National Science Foundation under CCF award 1910800. 

\nocite{HASSAN2021105008,ChauwinPRA2019,Herr2011,WalkerAPL2021,Edwards2022}
\bibliographystyle{IEEEtran}
\bibliography{IEEEabrv,main}

\end{document}